\documentclass[11pt]{article}
\pdfoutput=1
\usepackage{graphicx}
\usepackage{epsfig}
\usepackage{amsfonts}
\usepackage{amscd}
\usepackage{latexsym}
\usepackage{amsmath,amssymb}
\usepackage{verbatim}
\usepackage{setspace}
\usepackage{color}
\usepackage{fancyhdr}
\usepackage{cite}
\usepackage{hyperref}
\usepackage{tensor}
\usepackage{xfrac}
\usepackage{comment}

\usepackage[textheight=9in, textwidth=6.5in, letterpaper]{geometry}
\def\half{{1\over 2}}
\numberwithin{equation}{section}

 \def\p{\partial}
 \def\bz{{\bar z}}
 
\def\0{{(0)}}
\def\1{{(1)}}
\def\2{{(2)}}

\def\co{{\cal O}}

\def\<{\langle }
\def\>{\rangle }

\def\G{\Gamma }

\def \eps {\epsilon}
\def\bz{{\bar z}}

\def\p{\partial}

\def\CS {{\cal S}}

\def\CS {{\cal S }}

\def\slr{$SL(2,{\mathbb R})_R$}

\newcommand{\bea}{\begin{eqnarray}}
\newcommand{\eea}{\end{eqnarray}}
\newcommand{\be}{\begin{equation}}
\newcommand{\ee}{\end{equation}}
\newcommand{\ba}{\begin{aligned}}
\newcommand{\ea}{\end{aligned}}

\def\be{\begin{equation}}
\def\ee{\end{equation}}
\def\beq{\be\begin{array}{c}}
\def\eeq{\end{array}\ee}

   \makeatletter
  \let\over=\@@over \let\overwithdelims=\@@overwithdelims
  \let\atop=\@@atop \let\atopwithdelims=\@@atopwithdelims
  \let\above=\@@above \let\abovewithdelims=\@@abovewithdelims
\renewcommand\section{\@startsection {section}{1}{\z@}%
                                   {-3.5ex \@plus -1ex \@minus -.2ex}
                                   {2.3ex \@plus.2ex}%
                                   {\normalfont\large\bfseries}}

\renewcommand\subsection{\@startsection{subsection}{2}{\z@}%
                                     {-3.25ex\@plus -1ex \@minus -.2ex}%
                                     {1.5ex \@plus .2ex}%
                                     {\normalfont\bfseries}}

\usepackage{tikz}
\usepackage{slashed}
\usetikzlibrary{calc}   
\usetikzlibrary{topaths}
\usetikzlibrary{decorations}
\usetikzlibrary{decorations.pathmorphing}

{\cfoot{\thepage}}
\begin{document}
\onehalfspacing
\begin{titlepage}
\unitlength = 1mm~\\
\vskip 3cm
\begin{center}

{\LARGE{\textsc{Holographic Symmetry Algebras for\\ Gauge Theory and Gravity}}}

\vspace{0.8cm}
Alfredo Guevara{$^{*\dagger}$}, Elizabeth Himwich{$^{*}$}, Monica Pate{$^{*\dagger}$}, and Andrew Strominger{$^{*}$}\\
\vspace{1cm}

{$^*$\it  Center for the Fundamental Laws of Nature, Harvard University, Cambridge, MA 02138, USA} \\
{$^\dagger$\it Society of Fellows, Harvard University, Cambridge, MA 02138, USA}

\vspace{0.8cm}

\begin{abstract}
 All 4D gauge and gravitational theories in asymptotically flat spacetimes contain an infinite number of non-trivial symmetries. They can  be  succinctly characterized by generalized 2D currents acting  on the celestial sphere. A complete classification of these symmetries  and their algebras is an open problem. Here we construct two towers of such 2D currents from positive-helicity photons, gluons, or gravitons with integer conformal weights. These generate the symmetries associated to an infinite tower of conformally soft theorems.  The current algebra commutators are  explicitly  derived from the poles in the OPE coefficients, and found to comprise a rich closed subalgebra of the complete symmetry algebra.  
\end{abstract}

\end{center}

\end{titlepage}

\tableofcontents

\section{Introduction}

A central question in physics is: ``What are the fundamental  symmetries of nature?" One aspect of this question motivates the search for Beyond-the-Standard-Model physics and Unification. A second aspect, and the focus of the present work, is fully characterizing the non-trivial symmetries inherent in  the laws of physics that have already been verified experimentally: namely, General Relativity (GR) and the Standard Model. 

It is surprising that we still don't have a complete answer to this question, or even a precise formulation of the question itself.  GR has a diffeomorphism symmetry, but this is 
really a redundancy of description,  often referred to as  a ``trivial"  symmetry. Of greater interest are ``non-trivial" symmetries which have, by Noether's theorem, associated conservation laws with measurable consequences, such as conservation of linear/angular momentum or boost charge. One might have suspected that the Poincar\'e symmetries of Special Relativity -- which imply these conservation laws -- are the only non-trivial symmetries of GR (in the asymptotically flat context considered here). Famously, BMS \cite{Bondi:1962px,Sachs:1962wk} showed in 1962 that this could not be the case, and consequently that there is no limit in which General reduces to Special Relativity. Only very recently \cite{Strominger:2013jfa,He:2014laa}, using soft theorems from quantum field theory, was it shown that there are an infinite number of non-trivial symmetries of GR with associated conserved charges. These comprise a subgroup of the symmetries considered by BMS and can be measured using the gravitational memory effect \cite{Strominger:2014pwa}. However, various developments \cite{Barnich:2011mi,Cachazo:2014fwa,Kapec:2014opa,Campiglia:2014yka,Campiglia:2015yka,Choi:2019sjs,Godazgar:2018qpq,Kol:2019nkc,Himwich:2020rro} have made it clear that these are not {\it all} of the non-trivial symmetries of GR. Currently, there is not even a proposal for a complete classification of the non-trivial symmetries of nature! For QED the situation is similarly unresolved \cite{He:2014cra,Kapec:2015ena,Strominger:2015bla,Lysov:2014csa,Campiglia:2015qka,Campiglia:2016hvg,Himwich:2019dug,Nande:2017dba,Campiglia:2018dyi}. There is active research on this topic from a variety of viewpoints \cite{Strominger:2013lka,Dumitrescu:2015fej,Pasterski:2015zua,Kapec:2016jld,Campiglia:2016efb,Pasterski:2017ylz,Kapec:2017tkm,Choi:2017ylo,Donnay:2018neh,Ashtekar:2018lor,Campiglia:2017mua,Schreiber:2017jsr,Stieberger:2018edy,Stieberger:2018onx,Fan:2019emx,Fotopoulos:2019vac,Fotopoulos:2019tpe,Fan:2020xjj,Fotopoulos:2020bqj,Pate:2019mfs,Pate:2019lpp,Puhm:2019zbl,Adamo:2019ipt,Nandan:2019jas,Guevara:2019ypd,Law:2019glh,Albayrak:2020saa,Gonzalez:2020tpi,Casali:2020vuy,Casali:2020uvr,Banerjee:2020kaa,Banerjee:2020zlg,Banerjee:2020vnt,Campiglia:2020qvc,Ball:2019atb,Laddha:2020kvp,Donnay:2020guq,CPWinPrep}. 

There are many ways to characterize symmetries, which should ultimately all be equivalent. A familiar and traditional method  is the canonical construction of symmetry generators as  conserved charges that commute with the Hamiltonian or the $\CS$-matrix.\footnote{We consider here the  asymptotically flat approximation in which the  cosmological constant vanishes and there is an $\CS$-matrix.} Other methods directly derive relations among $\CS$-matrix elements, such as those given by soft theorems, or assume falloffs and perform an asymptotic symmetry analysis.  An especially powerful, recently-developed ``celestial" approach employs the holographic reformulation of the 4D $\CS$-matrix as a 2D conformal correlator on the celestial sphere at null infinity. In this approach, non-trivial symmetries correspond to generalized conformal currents on the celestial sphere. Their properties and algebra can be efficiently computed using the constraints of 2D conformal invariance. A further advantage  is that one largely avoids ambiguities associated to gauge choices, boundary terms, and falloff conditions. This paper takes a step towards classifying the non-trivial symmetries of nature within this approach. 

The symmetry-generating currents are of two types, arising from positive or negative helicity conformally soft massless particles. Whenever two opposite-helicity soft particles are scattered, the result can depend on the order of soft limits and a prescription of some kind is required to define the $\cal S$-matrix. In this paper we sidestep this important issue by considering only positive helicity currents\footnote{As noted in \cite{Banerjee:2020vnt}, this restriction arises automatically when considering the MHV sector, in which one helicity decouples.} and working in a  Vir$_L\otimes$\slr-invariant formalism.\footnote{Throughout this paper we treat left and right movers as independent on the celestial sphere, which means we effectively work in $(2,2)$ signature, $i.e.$ Klein space \cite{Atanasov:2021oyu}. The \slr\ here is the global subgroup of the Vir$_R$ superrotations.} Consistent with this restriction, this paper reports on a tower of higher-spin symmetry generators forming a closed generalized current-algebra sector of the celestial CFT$_2$. The symmetries we find are only a subgroup of all of the symmetries in gravity and gauge theory, but a large and interesting one. 

This infinite tower of symmetries is likely related to the infinite tower of soft theorems that have been discussed in the literature \cite{Li:2018gnc,Campiglia:2018dyi,Hamada:2018vrw}. We will show, however, that commutators of the leading, subleading, and (in gravity) subsubleading symmetries generate the whole tower, so the new symmetries here give no new constraints on the $\CS$-matrix.

In practice our discussion is largely for tree-level Einstein-Yang-Mills theory, but the methods are generally applicable and we anticipate that the algebra  persists in some form in the presence of quantum corrections and arbitrary higher-dimension operators  coming from UV physics. The results for gluons are affected if IR confinement occurs. The algebra is also affected by a short list of higher-dimension operators \cite{Laddha:2017vfh,Elvang:2016qvq} which deform the subleading soft theorems. 

The pure gluon algebra is derived in Section \ref{subsec:Gluons}. In 4D nonabelian gauge theory with group $G$, the leading soft theorem implies a standard closed 2D celestial $G$-current algebra \cite{Strominger:2013lka,He:2015zea,Fan:2019emx,Adamo:2019ipt,Pate:2019mfs}. The subleading soft theorem implies two further $G$-valued holomorphic  currents \cite{Lysov:2014csa,Pate:2019lpp,Banerjee:2020vnt}. We show that these form an \slr\ doublet. Commutators of two of these currents give yet further symmetry generators, which form an \slr\ triplet. Continuing in this manner, we construct an infinite tower of $G$-currents in finite-dimensional \slr\ representations and present their algebra. These currents have integral left$+$right conformal dimensions  $\Delta=1,0,-1,...$ where the operator product expansion is known to contain poles \cite{Pate:2019lpp,Guevara:2019ypd}. Spacetime translation invariance, which is not manifest in this presentation, combines these currents into a  representation of the Poincar\'e group. 

In Section \ref{subsec:Gravitons}, we consider the algebra of conformally soft positive helicity gravitons. Here the $\Delta=1$  \slr-doublet current generates supertranslations. The $\Delta=0$ \slr-triplet current closes with itself.\footnote{The shadow of this current is the Vir$_R$ stress tensor in \cite{Kapec:2016jld}.} At $\Delta=-1$ one encounters the \slr-quadruplet current associated to the subsubleading soft graviton theorem. These three currents then generate an infinite tower of currents, whose algebra we present. In Section \ref{subsec:GluonsGravitons} we couple gluons and gravitons and determine the resulting algebra. A directly analogous result applies to the coupling between photons and gravitons, although we do not include explicit formulae for that case.

In Section \ref{sec:amps}, we show explicitly that the OPEs we derive are encoded in four-gluon MHV scattering amplitudes, a result that can be extended to general multiplicity using the BCFW construction as described in \cite{Guevara:2019ypd}. Section \ref{sec:Summary} collects our results for the full \slr\ gluon and graviton algebra and we describe the straightforward application of these results to photons. Appendix \ref{app:Gluons} contains details of the gluon-gluon OPE calculation, which the graviton-graviton and gluon-graviton OPE calculations closely mimic. Appendix \ref{app:block} presents a compact, manifestly conformally covariant formula for the contribution to an OPE from a primary and all its $SL(2, \mathbb{R})_L \otimes SL(2, \mathbb{R})_R$ descendants. While this work was in progress \cite{Banerjee:2020zlg,Banerjee:2020vnt} appeared with overlapping results. 

\section{Gluons} \label{subsec:Gluons}

In this section, we find a class of positive helicity gluon operators $O^{a,+}_{\Delta}(z,\bar{z})$ at special conformal weights that generate symmetries of 
the $\CS$-matrix.\footnote{Following the conventions of \cite{Pate:2019lpp}, $O_{\Delta}^{a,\pm}$ denotes an outgoing conformal primary gluon of conformal (or boost) weight $\Delta=h+\bar h$, adjoint group index $a$, and helicity $s = \pm 1$. The group structure constants obey the Jacobi identity 
\begin{equation*} 
  f^{ab}{}_df^{dce} + f^{bc}{}_df^{dae} + f^{ca}{}_df^{dbe} = 0. 
\end{equation*} 
Details of the map from the momentum-space $\mathcal{S}$-matrix to celestial correlation functions are included in Section \ref{sec:amps}.} 
Our starting point is the leading OPE for two conformal primary, positive helicity gluons:\begin{equation} \label{eq:OOx}
  O^{a,+}_{\Delta_1}(z_1,\bar{z}_1)O^{b,+}_{\Delta_2}(z_2,\bar{z}_2) \sim  \frac{-i f^{ab}{}_c }{z_{12}} B(\Delta_1 - 1,\Delta_2 - 1) O^{c,+}_{\Delta_1 + \Delta_2 - 1}(z_2,\bar{z}_2), 
\end{equation}
where $z_{12}=z_1-z_2$ and $B(x,y)={\G(x)\G(y) \over \G(x+y)}$ is the Euler beta function. This expression was derived from a Mellin transform of the known soft collinear gluon singularities in \cite{Fan:2019emx} and from symmetries in \cite{Pate:2019lpp}. It is considerably less complicated than the general OPE \cite{Pate:2019lpp} that involves both helicities of gluons as well as gravitons.\footnote{Corrections to this OPE from higher-dimension operators are suppressed  by positive powers of $z_{12}$ or $\bz_{12}$ and would largely not affect the following analysis \cite{Pate:2019lpp}. The one interesting exception is $F^3$, which produces a potentially relevant term proportional 
$\frac{\bz_{12}^2}{z_{12}} O^{a,-}_{\Delta_1+\Delta_2+1}$.}
A salient feature of \eqref{eq:OOx} is the infinite sequence  of poles in the OPE coefficient  at integral conformal weights  $\Delta_1=k=1,0,-1,...$.  As we shall see, the fact that the weights are typically negative leads to some interesting and perhaps unfamiliar structures.  The highest weight case with $\Delta=1$ corresponds to the conformally soft gluon current  and turns out to be holomorphic. Contour integrals of this current on the celestial sphere generate the large gauge symmetries of the $\CS$-matrix \cite{He:2015zea}. The case $\Delta=0$ is the subleading soft current, which lies in an \slr \ doublet with negative  $\bar h=-{1 \over 2}$.\footnote{The zero mode of the leading $\Delta=1$ current generates global  color rotations, which are a global symmetry of the vacuum. This is the only unbroken global symmetry: all higher modes of the leading current are spontaneously broken, as well as the all modes of the further subleading currents.} Contour integrals of these currents generate subleading soft symmetries \cite{Lysov:2014csa,Himwich:2019dug}. The properties of the rest of the operators with integral $\Delta<0$, which generate an infinite tower of soft theorems, have not been well understood. Here we will construct the  2D algebra of these currents and find that it is closed. This tower is likely not yet the complete celestial symmetry algebra, which may  also include Goldstone currents \cite{Nande:2017dba,Donnay:2018neh,Himwich:2020rro,Donnay:2020guq} and negative helicity gluon and graviton generators or their shadows. However, it is an interesting and  non-trivial subalgebra thereof.

A description of the algebra that is manifestly covariant under the full Vir$_L\otimes$Vir$_R$ 2D conformal group is challenging. Here we cut the problem down to manageable size by presenting the  algebra  in a manifestly Vir$_L\otimes$\slr-covariant  form.  This is natural because the positive helicity symmetry currents fall into {\it finite} $(2-k)$-dimensional \slr \ representations with $\bar h ={k-1 \over 2}$ and highest (lowest) weights  ${1-k \over 2}$ (${k-1 \over 2}$).\footnote{This is equivalent to the null states found in \cite{Banerjee:2020vnt}.} The tower starts with (and we will see is generated by) the leading $k=1$ and subleading $k=0$ soft currents, which are a singlet and doublet, respectively. Mode-expanding on the right 
\begin{equation}
{O}_{k}^{a,+}(z,\bar{z}) = \sum_n\frac{{O}^{a,+}_{k,n}(z)}{\bar{z}^{n+{k-1 \over 2}}},
\end{equation}
 conformally soft currents are  defined by\footnote{Outside the specified range of $n$, the \slr-invariant norm vanishes. Such operators may still have contact interactions but in this paper operators are always at distinct  points.} 
 \begin{equation} \label{eq:Ropsx}
  R^{k,a}_n(z)
     := \lim_{\varepsilon \to 0} \varepsilon O_{k + \varepsilon,n}^{a,+}(z), ~~~~~ \ \ \ \ k = 1, 0, -1, -2, \ldots, ~~~~~~\quad {k-1 \over 2}\le n\le {1-k \over 2}.
\end{equation}
 For future notational convenience we also define 
 \begin{equation} \label{eq:Ropsxb}
  R^{k,a}(z,\bar{z})
    = \sum_{n={k-1 \over 2}}^{1-k \over 2}\frac{R^{k,a}_{n}(z)}{\bz^{n+\frac{k-1}{2}}} ,
\end{equation}
which has weights 
\begin{equation} \label{eq:Rweights}
  (h, \bar{h}) = \left(\frac{k+1}{2},  \frac{k-1}{2}\right). 
\end{equation}
These values of conformal weights $\Delta=k$  include all the conformally soft poles encountered in the OPE \eqref{eq:OOx}.  The factor of $\varepsilon$ incorporated in \eqref{eq:Ropsx} is needed to cancel these poles, leading to finite OPEs for the rescaled $R^{k,a}$.

Using $e.g.$ conformal blocks (see OPE blocks from \cite{Czech:2016xec}, reviewed in App.~\ref{app:block}) to derive the contribution from \slr\ descendants, the OPE \eqref{eq:OOx} has the further expansion in $\bz_{12}$
\begin{equation} \label{eq:OOdescendants}
  O^{a,+}_{\Delta_1}(z_1,\bar{z}_1)O^{b,+}_{\Delta_2}(z_2,\bar{z}_2) \sim \frac{-i f^{ab}{}_c }{z_{12}} \sum_{n = 0}^{\infty}  B(\Delta_1 - 1 + n, \Delta_2 - 1) \frac{(\bar{z}_{12})^n}{n!}\bar{\partial}^n O^{c,+}_{\Delta_1 + \Delta_2 - 1}(z_2,\bar{z}_2).
\end{equation}
The OPE of the conformally soft gluon operators \eqref{eq:Ropsxb}  then becomes
\begin{equation} \label{eq:RR}
 R^{k,a}(z_1,\bar{z}_1) R^{l,b}(z_2,\bar{z}_2) \sim \frac{-i f^{ab}{}_c }{z_{12}} \sum_{n = 0}^{1-k} 
 {2-k-l-n \choose 1-l}
 {(\bar{z}_{12})^n\over n!}\bar{\partial}^nR^{k+l -1,c}(z_2,\bar{z}_2),
\end{equation}
where in truncating the sum we make use of $\bar\p^{2-k}  R^{k,a}=0$. We can consider derivatives 
\begin{equation} \label{eq:dRdR}
  \bar{\partial}^p R^{k,a}(z_1,\bar{z}_1) \bar{\partial}^q R^{l,b}(z_2,\bar{z}_2) \sim  
  {-i f^{ab}{}_c\over z_{12}}
  {2-k-l-p-q \choose 1-l-q}\bar{\partial}^{q+p}R^{k+l -1,c}(z_2,\bar{z}_2), 
\end{equation}
where $0 \leq p < 2-k$ and $0 \leq q < 2 - l$ and we now include only leading order terms in both $z_{12}$ and $\bz_{12}$. Defining the commutator for holomorphic objects\footnote{Note that this is a 2D celestial commutator on a 1D circle, not to be mistaken for a 4D commutator on a 3D slice.}
\be  \left[A,B\right](z) = \oint_z \frac{dw}{2\pi i} A(w)B(z),\ee
and mode-expanding on the right, \eqref{eq:dRdR} can be reexpressed 
\begin{equation} \label{eq:RRcommutator}
\left[R^{k,a}_{n},R^{l,b}_{n'}\right] = -i f^{ab}{}_c { \frac{1-k}{2} -n + \frac{1-l}{2} - n' \choose 
\frac{1-k}{2} - n}{ \frac{1-k}{2} +n + \frac{1-l}{2} + n' \choose 
\frac{1-k}{2} +n} R^{k+l-1,c}_{n+n'}.
\end{equation}
This is the general conformally soft gluon algebra. 
For details of the derivation, see Appendix \ref{app:Gluons}. One may verify directly that this expression satisfies the Jacobi identity and is translation covariant.\footnote{Translation covariance follows from the Jacobi identity with conformally soft gravitons which are studied in subsequent sections. See for example \eqref{eq:HRope}.}  

The lowest-weight \slr\ element  $R^{k,a}_{k-1 \over 2}(z)$ for each $k$ is annihilated by $\bar L_{-1}=i\bar \p$, and hence is holomorphic. These operators have an especially simple chiral algebra. Define 
\begin{equation}\label{kil}
  \begin{aligned}
    \widehat{R}^{k,a}(z) &:= \bar{\partial}^{  1-k}R^{k,a}(z,\bar{z})=(1-k)!R^{k,a}_{k-1 \over 2}(z).  \end{aligned}
\end{equation}
Setting $p=1-k$ and $q=1-l$  in \eqref{eq:dRdR}, we obtain the relation
\begin{equation} \label{eq:ddR}
 \widehat R^{k,a}(z_1) \widehat  R^{l,b}(z_2) \sim \frac{-i f^{ab}{}_c }{z_{12}}  \widehat R^{k+l -1,c}(z_2).
\end{equation}
Equivalently,
\begin{equation} \label{eq:ddcR} \left[ \widehat R^{k,a}, \widehat  R^{l,b} \right] = -i f^{ab}{}_c \widehat R^{k+l -1,c}.
\end{equation}
Hence the algebra of the \slr\ lowest-weight states considerably simplifies, as does that of the \slr\ highest-weight states. Most of the complexity in \eqref{eq:RRcommutator} arises from the \slr\ mode level structure. 

\section{Gravitons} \label{subsec:Gravitons}

The analysis for positive-helicity gravitons $G_{\Delta}^{+}(z,\bar{z})$ is analogous to that for gluons.\footnote{Again following \cite{Pate:2019lpp}, $G_{\Delta}^{\pm}(z,\bar{z})$ denotes an outgoing conformal primary graviton of weight $\Delta$ and helicity $s = \pm 2$.} Define a family of conformally soft positive-helicity gravitons
\begin{equation} \label{eq:Hops}
H^{k} = \lim_{\varepsilon \to 0} \varepsilon G_{k + \varepsilon}^{+} , \ \ \ \ k = 2,1, 0, -1,  \ldots, 
\end{equation}
with weights
\begin{equation} \label{eq:Hweights}
 (h,\bar h)= \left(\frac{k+2}{2},  \frac{k-2}{2}\right),
\end{equation}
and a consistently truncated antiholomorphic mode expansion,
\begin{equation} \label{eq:Hmodes}
\begin{aligned}
  H^{k}(z,\bar{z}) &= \sum_{n = \frac{k-2}{2}}^{\frac{2-k}{2}} \frac{H^{k}_{n}(z)}{\bar{z}^{n + \frac{k-2}{2}}}.   \end{aligned}
\end{equation}
The $k=1$ term generates supertranslations. Expanding $H^1_n(z)=\sum_mH^1_{m,n}z^{-m-3/2}$, the four modes $H^1_{\pm \half,\pm \half}$ generate the four global translations. 

The OPE of two conformal primary gravitons of arbitrary weight \cite{Pate:2019lpp}, including antiholomorphic descendants, is 
\begin{equation} \label{eq:GG}
   G_{\Delta_1}^{+}(z_1,\bar{z}_1)G_{\Delta_2}^{+}(z_2,\bar{z}_2) \sim -\frac{\kappa}{2} \frac{1}{z_{12}} \sum_{n=0}^{\infty} B(\Delta_1 - 1 +n , \Delta_2 -1) \frac{(\bar{z}_{12})^{n+1}}{n!} \bar{\partial}^{n}G^{+}_{\Delta_1 + \Delta_2}(z_2,\bar{z}_2),
\end{equation}
with $\kappa = \sqrt{32\pi G}$. The OPE of conformally soft gravitons \eqref{eq:Hops} becomes 
\begin{equation}
H^{k}(z_1,\bar{z}_1) H^{l}(z_2,\bar{z}_2) \sim -\frac{\kappa}{2} \frac{1}{z_{12}} \sum_{n=0}^{1-k}  {2-k-l-n\choose 1-l} \frac{(\bar{z}_{12})^{n+1}}{n!} \bar{\partial}^{n}H^{k + l}(z_2,\bar{z}_2).
\end{equation}
After some algebra (analogous to the gluon case in App.~\ref{app:Gluons}) one finds the current commutators 
\begin{equation}\label{sxa}
\left[H^{k}_{m},H^{l}_{n}\right] = -\frac{\kappa}{2} \left[ n(2-k) - m(2-l) \right]\frac{(\frac{2-k}{2} -m + \frac{2-l}{2} - n -1)!}{( \frac{2-k}{2} - m)!(\frac{2-l}{2} - n)!} \frac{(\frac{2-k}{2} + m + \frac{2-l}{2} + n - 1)!}{(\frac{2-k}{2}+m)!(\frac{2-l}{2}+n)!} H^{k+l}_{m+n}.
\end{equation}
As in the gauge theory case, we check that this commutator obeys the Jacobi identity with three $H$ operators.  These imply the $H$ operators obey the Jacobi identity with $\bar{L}_0,\bar{L}_{\pm 1}$, and that it is translation covariant. 

For the case of gluons, there is a closed subalgebra with $k=1$. For gravitons, the closed subalgebra has $k=0$. Defining 
\be J_1=-{2\over \kappa}H^0_1,~~~J_0={1\over \kappa}H^0_0,~~J_{-1}=-{2\over \kappa}H^0_{-1}, \ee one finds from \eqref{sxa} the current algebra
\begin{equation}\label{snxa}
\left[J_{m},J_{n}\right] = (m-n)J_{m+n}. 
\end{equation}
The zero modes of $J_m$  generate self-dual Lorentz transformations in Klein space.

\section{Gluons and Gravitons} \label{subsec:GluonsGravitons}

As in the previous sections, we begin with the OPE of conformally soft gluons and gravitons of arbitrary weight derived in \cite{Pate:2019lpp}, and include antiholomorphic descendants:
\begin{equation}
 G_{\Delta_1}^{+}(z_1,\bar{z}_1)O_{\Delta_2}^{a,+}(z_2, \bar{z}_2) \sim - \frac{\kappa}{2}\frac{1}{z_{12}} \sum_{n = 0}^{\infty} B(\Delta_1 +n - 1, \Delta_2)\frac{\bar{z}_{12}^{n+1}}{n!} \bar{\partial}^{n} O^{a,+}_{\Delta_1 + \Delta_2}(z_2,\bar{z}_2).
\end{equation}
In terms of the conformally soft gluons and gravitons defined above, the OPE becomes 
\begin{equation}
    H^{k}(z_1,\bar{z}_1) R^{l,a}(z_2,\bar{z}_2) \sim - \frac{\kappa}{2}\frac{1}{z_{12}} \sum_{n = 0}^{1-k} \frac{(1-k-l-n)!}{(1-k-n)!(-l)!} \frac{\bar{z}_{12}^{n+1}}{n!} \bar{\partial}^{n}R^{k+l,a}(z_2,\bar{z}_2), \\
\end{equation}
where we again make use of the finite  \slr\ representation to truncate the sum. After a calculation analogous to that in previous sections, we derive the commutator 
\begin{equation} \label{eq:HRope}
\left[H^{k}_{m},R^{l,a}_{n}\right] = -\frac{\kappa}{2} \left[ n(2-k) - m(1-l) \right]\frac{(\frac{2-k}{2} -m + \frac{1-l}{2} - n - 1)!}{( \frac{2-k}{2} - m)!(\frac{1-l}{2} - n)!} \frac{(\frac{2-k}{2} + m + \frac{1-l}{2} + n- 1)!}{(\frac{2-k}{2}+m)!(\frac{1-l}{2}+n)!} R^{k+l,a}_{m+n}.
\end{equation}

\section{OPE from Scattering Amplitudes} \label{sec:amps}

In order to make contact with other work on  scattering amplitudes, we will show how the previous OPEs including all antiholomorphic descendants emerge when the celestial primaries are identified with scattering states in MHV tree amplitudes (see also {\cite{Banerjee:2020kaa,Ebert:2020nqf}}). In momentum space, massless $n$-particle amplitudes  $\textbf{A}(\epsilon_i\omega_i, z_i)$ are labeled by energies $\omega_i$ and points $z_i$ on the celestial sphere. This follows from a parametrization of massless momenta
\begin{equation}
p_k^{\mu} = \frac{\epsilon_k\omega_k}{\sqrt{2}} \left(1 + z_k\bar{z}_k, z_k + \bar{z}_k, - i(z_k - \bar{z}_k), 1 - z_k\bar{z}_k\right),
\end{equation}
with $\epsilon_k = \pm 1$ for outgoing and incoming momenta, respectively. Celestial amplitudes $\boldsymbol{\mathcal{A}}(\Delta_i, z_i)$ are defined by transforming to Mellin space (see $e.g.$   \cite{Pasterski:2016qvg,Pasterski:2017kqt}) \begin{equation}\label{eq:mellin}
\boldsymbol{\mathcal{A}}_{s_j}(\Delta_j, z_j) = \left(\prod_{i=1}^{n} \int_0^{\infty} \frac{d\omega_i}{\omega_i} \omega_i^{\Delta_i} \right) \textbf{A}_{s_i}(\epsilon_i\omega_i, z_i), 
\end{equation}
where $s_i$ are helicity labels.\footnote{Mellin amplitudes converge only for certain $\Delta_i$ but can be defined generally by analytic continuation \cite{Arkani-Hamed:2020gyp,Donnay:2020guq}.} Celestial amplitudes are naturally interpreted as correlation functions of $n$ weight $\left(h_i, \bar{h}_i\right) = \left(\frac{\Delta_i + s_i}{2}, \frac{\Delta_i - s_i}{2}\right)$ conformal primary operators on the celestial sphere:
\begin{equation}\label{eq:amptocorr}
\boldsymbol{\mathcal{A}}_{s_i}(\Delta_i, z_i) \rightarrow \langle \mathcal{O}_{\Delta_1}^{s_1}(z_1,\bar{z}_1) \cdots \mathcal{O}_{\Delta_n}^{s_n}(z_n,\bar{z}_n)\rangle.
\end{equation}
Let us consider first the case of gluons with $s_i = \pm 1$. Nicely, the full OPE \eqref{eq:OOdescendants} is encoded even in the simplest non-trivial scattering process, namely $n=4$ gluons.
We start from 
\begin{align}
\mathbf{A}_{--++}^{abcd}(\{\eps_1\omega_{1},z_{1},\bar{z}_{1}\},\ldots,\{\eps_4\omega_{4},z_{4},\bar{z}_{4}\}){=}&\delta^{4}\left(\sum_{i=1}^{4}\epsilon_{i}\lambda_{i}^{\alpha}\tilde{\lambda}_{i}^{\dot{\alpha}}\right)\times \nonumber \\
&\left[f^{ad}{}_ef^{cbe}\frac{\langle12\rangle^{3}}{\langle23\rangle\langle34\rangle\langle41\rangle}{+}f^{ac}{}_ef^{dbe}\frac{\langle12\rangle^{3}}{\langle24\rangle\langle43\rangle\langle31\rangle}\right]\,,\label{eq:4ptmhv}
\end{align}
with null momenta written as $p_{\alpha\dot{\alpha}} = \epsilon \lambda_{\alpha}\tilde{\lambda}_{\dot{\alpha}}$ using 
\begin{equation}
\lambda_{i}=\sqrt{\omega_{i}}(1\,\,z_{i})\,\,\,,\,\,\tilde{\lambda}_{i}=\sqrt{\omega_{i}}\left(\begin{array}{c}
1\\
\bar{z}_{i}
\end{array}\right)\,,\label{eq:lamp}
\end{equation}
where $\langle ij\rangle=\epsilon_{i}\epsilon_{j}\lambda_{i}^{\alpha}\lambda_{j}^{\beta}\epsilon_{\alpha\beta}\,$. We will consider the holomorphic collinear limit for positive-helicity gluons $3,4$, which we take as outgoing $\epsilon_{3}=\epsilon_{4}=+1$. To study only the antiholomorphic descendants, using the parametrization (\ref{eq:lamp}), we can extract the leading order behavior as $z_{3}\to z_{4}$ while keeping the exact dependence on $\bar{z}_{34}$:  
\begin{align}
\delta^{4}\left(\sum_{i=1}^{4}\epsilon_{i}\lambda_{i}^{\alpha}\tilde{\lambda}_{i}^{\dot{\alpha}}\right) & =\frac{1}{ \omega_{1}\omega_{2}z_{12}^{2}}\delta\left(\epsilon_{1}\omega_{1}{-}\frac{z_{24}}{z_{12}}(\omega_{3}+\omega_{4})\right)\delta\left(\epsilon_{2}\omega_{2}{-}\frac{z_{41}}{z_{12}}(\omega_{3}+\omega_{4})\right)\delta\left(\bar{z}_{14}{-}\epsilon_{1}\frac{\omega_{3}}{\omega_{1}}\frac{z_{24}}{z_{12}}\bar{z}_{34}\right)\nonumber \\
 & \,\,\times\delta\left(\bar{z}_{24}-\epsilon_{2}\frac{\omega_{3}}{\omega_{2}}\frac{z_{41}}{z_{12}}\bar{z}_{34}\right)+\mathcal{O}(z_{34})\,.
\end{align}
Introducing $\omega_{3}=t\omega,\omega_{4}=(1-t)\omega$ with $0\leq t\leq1$, this becomes 
\begin{align}
\frac{1}{ \omega_{1}\omega_{2}z_{12}^{2}}\delta\left(\epsilon_{1}\omega_{1}-\frac{z_{24}}{z_{12}}\omega\right)\delta\left(\epsilon_{2}\omega_{2}-\frac{z_{41}}{z_{12}}\omega\right)\delta\left(\bar{z}_{14}-t\bar{z}_{34}\right)\delta\left(\bar{z}_{24}-t\bar{z}_{34}\right)\nonumber \\
=\left.\delta^{4}\left(\sum_{i=1,2,4}\epsilon_{i}\lambda_{i}^{\alpha}\tilde{\lambda}_{i}^{\dot{\alpha}}\right)\right|_{\bar{z}_{4}\to\bar{z}_{4}+t\bar{z}_{34},~\omega_4\to \omega},
\end{align}
$i.e.$ at leading order in $z_{34}$, the $n=4$ momentum conservation condition can be written as the $n=3$ condition together with the deformation
\begin{equation}
\bar{z}_{4}\to\bar{z}_{4}+t\bar{z}_{34}\,.\label{eq:def4}
\end{equation}
This deformation extends trivially to the full amplitude (\ref{eq:4ptmhv}) since the remaining factor does not depend on the antiholomorphic coordinates. Indeed, at leading order in $z_{34}\sim\langle34\rangle$ the stripped part of (\ref{eq:4ptmhv}) becomes
\begin{equation}
f^{ad}{}_ef^{cbe}\frac{\langle12\rangle^{3}}{\langle23\rangle\langle34\rangle\langle41\rangle}+f^{ac}{}_ef^{dbe}\frac{\langle12\rangle^{3}}{\langle24\rangle\langle43\rangle\langle31\rangle}\to\frac{f^{cd}{}_e}{z_{34}\omega t(1-t)}\times\left( f^{abe}\frac{\langle12\rangle^{3}}{\langle24\rangle\langle41\rangle}\right)_{\omega_4\to \omega},
\end{equation}
and we can relate the full $n=4$ amplitude to an $n=3$ amplitude:
\be\label{eq:4to3pt}
		\begin{split}
		{\bf A}^{abcd}_{--++}(\{ &\eps_1 \omega_1, z_1, \bz_1\}, \{ \eps_2 \omega_2, z_2, \bz_2\},\{  \omega_3, z_3, \bz_3\},\{  \omega_4, z_4, \bz_4\})\\
			 &= \frac{-i f^{cd}{}_e}{z_{34} \omega t (1-t) } {\bf A}^{abe}_{--+}(\{ \eps_1 \omega_1, z_1, \bz_1\}, \{ \eps_2 \omega_2, z_2, \bz_2\},\{ \omega, z_4, \bz_4+ t\bz_{34}\})+ \co (z_{34}^0).
		\end{split}
	\ee
To translate the above statement into a celestial correlation function we employ \eqref{eq:mellin}-\eqref{eq:amptocorr} and find 
\begin{align}\label{eq:4ptcorrmell}
\langle O_{\Delta_{1}}^{a,-}&\cdots O_{\Delta_{4}}^{d,+}\rangle \nonumber \\ &=\int d\omega_{1} \ \omega_{1}^{\Delta_{1}-1}d\omega_{2} \ \omega_{2}^{\Delta_{2}-1}d\omega \ \omega^{\Delta_{3}+\Delta_{4}-1}dt \ t^{\Delta_{3}-1}(1-t)^{\Delta_{4}-1}\mathbf{A}_{--++}^{abcd}\nonumber \\
 & \to\frac{-if^{cd}{}_e}{z_{34}}\int dt \ t^{\Delta_{3}-2}(1-t)^{\Delta_{4}-2}\nonumber \\
 & \,\,\,{\times}\int d\omega_{1} \ \omega_{1}^{\Delta_{1}-1}d\omega_{2} \ \omega_{2}^{\Delta_{2}-1}d\omega \ \omega^{\Delta_{3}+\Delta_{4}-2}\mathbf{A}_{--+}^{abe}(\{-\omega_{1},z_{1},\bar{z}_{1}\},\{-\omega_{2},z_{2},\bar{z}_{2}\},\{\omega,z_{4},\bar{z}_{4}{+}t\bar{z}_{34}\})\nonumber \\
 & =\frac{-if^{cd}{}_e}{z_{34}}\int dt \ t^{\Delta_{3}-2}(1-t)^{\Delta_{4}-2}\langle O_{\Delta_{1}}^{a,-}(z_{1},\bar{z}_{1})O_{\Delta_{2}}^{b,-}(z_{2},\bar{z}_{2})O_{\Delta_{3}+\Delta_{4}-1}^{e,+}(z_{4} ,\bar{z}_{4}+t\bar{z}_{34})\rangle\,,
\end{align} 
where gluons $1,2$ are incoming and $3,4$ are outgoing. The OPE of two positive helicity gluons follows:
\begin{equation}\label{eq:OOblock}
O_{\Delta_{3}}^{c,+}(z_{3},\bar{z}_{3})O_{\Delta_{4}}^{d,+}(z_{4},\bar{z}_{4})\sim\frac{-if^{cd}{}_e}{z_{34}}\int dt \ t^{\Delta_{3}-2}(1-t)^{\Delta_{4}-2}O_{\Delta_{3}+\Delta_{4}-1}^{e,+}(z_{4},\bar{z}_{4}+t\bar{z}_{34}).
\end{equation}
This expression corresponds to a conformal block including all antiholomorphic descendants. This connection is detailed in Appendix~\ref{app:block}. To see this explicitly we perform
a Taylor expansion in $\bar{z}_{34}$: 
\begin{align}
\nonumber O_{\Delta_{3}}^{c,+}(z_{3},\bar{z}_{3})O_{\Delta_{4}}^{d,+}(z_{4},\bar{z}_{4}) & \sim\frac{-if^{cd}{}_e}{z_{34}}\sum_{n=0}^{\infty}\int dt \ t^{\Delta_{3}-2+n}(1-t)^{\Delta_{4}-2}\frac{\bar{z}_{34}^{n}}{n!}\bar{\partial}^{n}O_{\Delta_{3}+\Delta_{4}-1}^{e,+}(z_{4},\bar{z}_{4})\\
 & \sim\frac{-if^{cd}{}_e}{z_{34}}\sum_{n=0}^{\infty}B(\Delta_{3}-1+n,\Delta_{4}-1)\frac{\bar{z}_{34}^{n}}{n!}\bar{\partial}^{n}O_{\Delta_{3}+\Delta_{4}-1}^{e,+}(z_{4},\bar{z}_{4})\,,
\end{align}
in agreement with \eqref{eq:OOdescendants}. For gravitons, analogous computations hold for the $n=4$ MHV celestial correlators. For instance, the four-graviton amplitude in \cite{Puhm:2019zbl} leads to the conformal block 
\begin{equation}
G_{\Delta_{3}}^{+}(z_{3},\bar{z}_{3})G_{\Delta_{4}}^{+}(z_{4},\bar{z}_{4})\sim -\frac{\kappa}{2}\frac{\bar{z}_{34}}{z_{34}}\int dt \ t^{\Delta_{3}-2}(1-t)^{\Delta_{4}-2}G_{\Delta_{3}+\Delta_{4}}^{+}(z_{4},\bar{z}_{4}+t\bar{z}_{34}),
\end{equation} 
which can be Taylor expanded to obtain \eqref{eq:GG}.

To see that the above gluon-gluon OPE is consistent with scattering
amplitudes at any multiplicity and general helicity configuration,
we generalize the expression \eqref{eq:4to3pt} as follows:
\begin{equation}
\begin{aligned}
\mathbf{A}_{\{s_{j}\}++}^{\{a_{j}\}cd}(\{\omega_{1},z_{1},\bar{z}_{1}\},\ldots,&\{\omega_{n+1},z_{n+1},\bar{z}_{n+1}\}) \\ &\sim\frac{-if^{cd}{}_e}{z_{n,n+1}\omega t(1-t)}\mathbf{A}_{\{s_{j}\}+}^{\{a_{j}\}e}(\{\omega_{1},z_{1},\bar{z}_{1}\},\ldots,\{\omega,z_{n},\bar{z}_{n}+t\bar{z}_{n+1,n}\})\label{eq:nton-1pt}
\end{aligned}
\end{equation}
where $s_{j}=\pm1,\,j=1,\ldots,n-1$ and we have set again $\omega_{n+1}=t\omega,\omega_{n}=(1-t)\omega$.
It is then straightforward, following the steps of eq. \eqref{eq:4ptcorrmell},
to show that the OPE \eqref{eq:OOblock} follows.

The formula \eqref{eq:nton-1pt} can be proven from the BCFW recurrence
relations of gluon scattering amplitudes, following the construction
of \cite{Guevara:2019ypd} (see also references therein). The $n+1$-point
scattering amplitude obtained by attaching a positive helicity gluon of coordinates $\{\omega_s,z_s,\bar{z}_s\}$ is written as
\begin{align}
\mathbf{A}_{\{s_{j}\}+}^{\{a_{j}\}a_{s}}&(\{\omega_{1},z_{1},\bar{z}_{1}\},\ldots,\{\omega_{s},z_{s},\bar{z}_{s}\}) \nonumber\\
& \sim\sum_{i=2}^{n}-if^{a_{s}a_{i}}{}_b\frac{z_{1i}}{z_{1s}z_{si}\omega_{s}}(1+\alpha_{i})^{s_{i}}(1+\beta_{i})^{s_{1}}\times\nonumber \\
 & \ \ \ \ \ \   \,\mathbf{A}_{\{s_{j}\}}^{\{a_{j},a_{i}\to b\}}(\{(1+\beta_{i})\omega_{1},z_{1},\frac{\bar{z}_{1}+\beta_{i}\bar{z}_{s}}{1+\beta_{i}}\},\ldots,\{(1+\alpha_{i})\omega_{i},z_{i},\frac{\bar{z}_{i}+\alpha_{i}\bar{z}_{s}}{1+\alpha_{i}}\},\ldots)\label{eq:bcfwn1}
\end{align}
where $\alpha_{i}=\frac{\omega_{s}z_{1s}}{\omega_{i}z_{1i}},\beta_{i}=\frac{\omega_{s}z_{is}}{\omega_{1}z_{i1}}$
and we have neglected, following \cite{Guevara:2019ypd}, multiparticle
factorizations which are regular in the OPE limit. Let us again consider
the leading behaviour as $z_{n}\to z_{s}$, keeping $\bar{z}_{ns}$
finite. For this we need only keep the term $i=n$ in the above sum and set $s_{n}=+1$ for the positive helicity gluon. Introducing $\omega_{s}=t\omega,\omega_{n}=(1-t)\omega$
as anticipated we find $\alpha_{n}\to\frac{t}{1-t},\beta_{n}\to0$
and the leading behaviour of \eqref{eq:bcfwn1} becomes precisely
\eqref{eq:nton-1pt}.

\section{Summary} \label{sec:Summary}

We collect the results for the conformally soft algebra derived in the previous sections:
\begin{equation}
\begin{aligned} 
\left[R^{k,a}_{n},R^{l,b}_{n'}\right] &= -i f^{ab}{}_c \frac{( \frac{1-k}{2} -n + \frac{1-l}{2} - n')!}{(\frac{1-k}{2} - n)!( \frac{1-l}{2} - n')!} \frac{(\frac{1-k}{2} + n + \frac{1-l}{2} +n')!}{(\frac{1-k}{2}+n)!(\frac{1-l}{2}+n')!} R^{k+l-1,c}_{n+n'}, \\
\left[H^{k}_{n},H^{l}_{n'}\right] &= -\frac{\kappa}{2} \left[ n'(2-k) - n(2-l) \right]\frac{(\frac{2-k}{2} -n + \frac{2-l}{2} - n' -1)!}{( \frac{2-k}{2} - n)!(\frac{2-l}{2} - n')!} \frac{(\frac{2-k}{2} + n + \frac{2-l}{2} + n' - 1)!}{(\frac{2-k}{2}+n)!(\frac{2-l}{2}+n')!} H^{k+l}_{n+n'}, \\
\left[H^{k}_{n},R^{l,a}_{n'}\right] &= -\frac{\kappa}{2} \left[ n'(2-k) - n(1-l) \right]\frac{(\frac{2-k}{2} -n + \frac{1-l}{2} - n' - 1)!}{( \frac{2-k}{2} - n)!(\frac{1-l}{2} - n')!} \frac{(\frac{2-k}{2} + n + \frac{1-l}{2} + n' - 1)!}{(\frac{2-k}{2}+n)!(\frac{1-l}{2}+n')!} R^{k+l,a}_{n+n'}.
\end{aligned}
\end{equation}
These results readily extend to photons.  In particular, one can construct generalized conformally soft photon currents that are directly analogous to the generalized conformally soft gluon currents.  Unlike gluons, which carry color, photons do not carry electric charge and as a result their generalized currents commute with one another.  However, since photons couple to gravitons, the photon currents obey commutation relations with the graviton currents that are of the same form as the gluon-graviton commutation relations.

\section*{Acknowledgements}

We are grateful to Alex Atanasov, Adam Ball, Dan Kapec, Walker Melton, and Ana Raclariu for useful conversations.  This work was supported by DOE grant de-sc/0007870.  AG and MP are supported by Junior Fellowships at the Harvard Society of Fellows.

\appendix

\section{Gluon OPE Calculation} \label{app:Gluons}

The OPE of generalized soft gluon operators $R^{k,a}$ is derived from the OPE of conformal primary gluons 
\be \label{ope_original}
    O^{a,+}_{\Delta_1}(z_1,\bar{z}_1)O^{b,+}_{\Delta_2}(z_2,\bar{z}_2) \sim \frac{-i f^{ab}{}_c }{z_{12}} \sum_{n = 0}^{\infty}  B(\Delta_1 - 1 + n, \Delta_2 - 1) \frac{\bar{z}_{12}^n}{n!} \partial_{\bz_2}^n O^{c,+}_{\Delta_1 + \Delta_2 - 1}(z_2,\bar{z}_2),
\ee
using the definition 
\be
    R^{k,a}(z, \bz) = \lim_{\varepsilon \to 0} \varepsilon O^{a,+}_{k+\varepsilon}(z,\bz), 
        \quad \quad \quad k = 1, 0, -1, -2, \cdots, 
\ee
with the mode expansion
\be \label{Rmode_exp}
    R^{k,a}(z, \bz) = \sum_{n = \frac{k-1}{2}}^{\frac{1-k}{2}}
        \frac{R^{k,a}_n(z)}{\bz^{n+\frac{k-1}{2}}}.
\ee
Taking the simultaneous limit of \eqref{ope_original}, we find
\be \label{R_OPE}
    \begin{split}
        R^{k,a}(z_1,\bz_1) R^{\ell, b}(z_2, \bz_2)
            & = \lim_{\varepsilon \to 0} \varepsilon O^{a,+}_{k+\varepsilon}(z_1, \bz_1)
            \varepsilon O^{b,+}_{\ell+\varepsilon}(z_2, \bz_2)\\
        & \sim \frac{-i f^{ab}{}_c }{z_{12}}
            \sum_{m = 0}^{1-k} \frac{1}{m!} \frac{(2-k-\ell-m)!}{(1-k-m)! (1-\ell)!}
                \bz_{12}^m \p^m_{\bz_{2}} R^{k+\ell-1, c}(z_2, \bz_2),
    \end{split}
\ee
where the truncation in the sum over $m$ follows from the mode expansion \eqref{Rmode_exp}. (Specifically, note that the powers of $\bz_1$ match on either side 
of the equation.)

To determine the algebra of modes $R^{k,a}_n$ from the OPE \eqref{R_OPE}, we first 
recall that modes are extracted from 
\be 
    R^{k, a}_n(z) = \oint \frac{d \bz}{ 2\pi i} \bz^{n + \frac{k-3}{2}}R^{k,a}(z, \bz).
\ee
Here $z$ and $\bz$ are treated independently, as, for example, is done in Section 6.1 of \cite{DiFrancesco:1997nk}. Then, the algebra of modes $R^{k,a}_n$ is obtained by taking the following contour integrals:
 \be \label{fullexp_commutator}
    \begin{split}
        \left[R^{k,a}_n, R^{\ell, b}_{n'}\right](z_2)
             = \oint_{|\bz_1|< \eps} \frac{d \bz_1}{2 \pi i } \bz_1^{n+\frac{k-3}{2}}
                \oint_{|\bz_2|< \eps} \frac{d \bz_2}{2\pi i}\bz_2^{n'+\frac{\ell-3}{2}}
                    \oint_{|z_{12}|< \eps} \frac{d z_1}{2 \pi i} ~R^{k,a}(z_1,\bz_1) R^{\ell, b}(z_2, \bz_2).
    \end{split}
 \ee
Note that since the OPE \eqref{R_OPE} is not singular in the antiholomorphic variables, the order in which contour integrals in antiholomorphic variables is taken does not matter. 
 Let's take the $\bz_1$ contour first. Substituting the OPE \eqref{R_OPE} into 
 the right-hand side of \eqref{fullexp_commutator}, performing the $z_1$ integral
 and using 
 \be
    \oint_{|\bz_1|< \eps} \frac{d \bz_1}{2 \pi i }~ \bz_1^{n+\frac{k-3}{2}} \bz_{12}^m
         =\left \{ 
         \begin{matrix}
         0, & 0\leq m < \frac{1-k}{2}-n\\
         \dfrac{m!}{(\frac{1-k}{2}-n)! (m+n+ \frac{k-1}{2})!}(-\bz_2)^{m+n+ \frac{k-1}{2}}, &  \frac{1-k}{2}-n\leq m \leq 1-k
         \end{matrix}\right. ,
 \ee
 to perform the $\bz_1$ integral, 
 \eqref{fullexp_commutator} becomes
 \be \label{fullexp_commutator2}
    \begin{split}
        \left[R^{k,a}_n, R^{\ell, b}_{n'}\right](z_2)
             &=-i f^{ab}{}_c    
            \sum_{m = \frac{1-k}{2}-n}^{1-k}   \frac{(2-k-\ell-m)!}{(1-k-m)! (1-\ell)!}
                 \frac{(-1)^{m+n+ \frac{k-1}{2}}}{(\frac{1-k}{2}-n)! (m+n+ \frac{k-1}{2})!} \\& \quad \quad \quad \quad \quad\quad \quad
                                            \quad \quad\times 
                \oint_{|\bz_2|< \eps} \frac{d \bz_2}{2\pi i}\bz_2^{m+n+ \frac{k-1}{2}+n'+\frac{\ell-3}{2}}  \p^m_{\bz_{2}} R^{k+\ell-1, c}(z_2, \bz_2) .
    \end{split}
 \ee
 To perform the remaining contour integral, we substitute $R^{k+\ell-1, c}(z_2, \bz_2)$ for its mode expansion
 \be
    \begin{split}
         \oint_{|\bz_2|< \eps} \frac{d \bz_2}{2\pi i}\bz_2^{m+n+ \frac{k-1}{2}+n'+\frac{\ell-3}{2}} &\p^m_{\bz_{2}} R^{k+\ell-1, c}(z_2, \bz_2) \\
         & = \oint_{|\bz_2|<\eps}  \frac{ d \bz_2}{2 \pi i}~\bz_2^{m+\frac{k-1}{2} +n+n'+ \frac{\ell-3}{2}}  
						\p^m_{\bz_2} \sum_{m'={k+\ell-2 \over 2}}^{2-k-\ell \over 2}\frac{R^{k+\ell-1,c}_{m'}(z_2)}{\bz_2^{m'+\frac{k+\ell-2}{2}}}\\ 
				& =      \frac{(\frac{1-k}{2}- n+ \frac{1-\ell}{2}-n')!}{(\frac{1-k}{2}- n+ \frac{1-\ell}{2}-n'-m)!}R^{k+\ell-1,c}_{ n+n' }(z_2).
    \end{split}
 \ee
 Finally, substituting this result back in \eqref{fullexp_commutator2}
 and performing the sum in $m$, we find
\be \label{fullexp_commutator_final}
    \begin{split}
        \left[R^{k,a}_n, R^{\ell, b}_{n'}\right](z_2)
             &=-i f^{ab}{}_c ~  \frac{( \frac{1-k}{2}-n+\frac{1-\ell}{2}-n')!}{( \frac{1-k}{2}-n)!(\frac{1-\ell}{2}-n')!}
					 	\frac{(\frac{1-k}{2}+n +\frac{1-\ell}{2}+n')!}{(\frac{1-k}{2}+n)!( \frac{1-\ell}{2}+n')!} R^{k+\ell-1,c}_{ n+n' }(z_2).
    \end{split}
 \ee 
 
 \section{The OPE Conformal Block} \label{app:block}

In this appendix we will derive the integral expression \eqref{eq:OOblock}
encoding the contribution from all $SL(2,\mathbb{R})_{R}$ descendants.
To do so we first start with the complete expression for $SL(2,\mathbb{R})_{L}\otimes SL(2,\mathbb{R})_{R}$
descendants as given in $e.g.$ Appendix B of \cite{Czech:2016xec}.\footnote{Recall we are working with independent left and right coordinates,
$i.e.$ $(2,2)$ signature \cite{Atanasov:2021oyu}.} Consider two primaries of weights $(h_{1},\bar{h}_{1}),(h_{2},\bar{h}_{2})$.
In the shadow representation, singling out the contribution from a
single primary $P$ and all its descendants leads to the expression 
\begin{equation}
\begin{aligned}
O_{(h_{1},\bar{h}_{1})}(z_{1,}\bar{z}_{1})&O_{(h_{2},\bar{h}_{2})}(z_{2},\bar{z}_{2}) \\
&\sim\mathcal{N}\int\frac{d^{2}z_{3}\,O_{(h_{P},\bar{h}_{P})}(z_{3},\bar{z}_{3})}{z_{12}^{h_{1}+h_{2}+h_{p}-1}z_{32}^{h_{2}-h_{1}-h_{P}+1}z_{13}^{h_{1}-h_{2}-h_{P}+1}\bar{z}_{12}^{\bar{h}_{1}+\bar{h}_{2}+\bar{h}_{P}-1}\bar{z}_{32}^{\bar{h}_{2}-\bar{h}_{1}-\bar{h}_{P}+1}\bar{z}_{13}^{\bar{h}_{1}-\bar{h}_{2}-\bar{h}_{P}+1}}\\
&=\mathcal{N}I_{O_{1}O_{2}}^{O_{3}},
\end{aligned}
\end{equation}
where $\mathcal{N}$ is a normalization constant that can be fixed
 by comparing leading order terms in the limit $z_{12}, \bz_{12} \to 0$. As coordinates $z,\bar{z}$ are independent
we will assume they both lie on the real projective line. If $z_{1}>z_{2},\bar{z}_{1}>\bar{z}_{2}$,
the KLT formula\footnote{To make connection with the standard four-point KLT formula as described
in \cite{Kawai:1985xq,Polchinski:1998rq}, it is convenient to set $z_1=1,z_2=0$ using $SL(2,\mathbb{R})$ covariance and expand  $O_{(h_{P},\bar{h}_{P})}(z_{3},\bar{z}_{3})$
in powers of $z_{3},\bar{z}_{3}$.} provides a factorization of the above integral into two disk integrals:  
\begin{equation}
\begin{aligned}
I_{O_{1}O_{2}}^{O_{3}}=2\sin(\pi(h_{2}-h_{1}+h_{P}))&\int_{z_{1}}^{\infty}\frac{dz_{3}}{z_{12}^{h_{1}+h_{2}+h_{p}-1}z_{32}^{h_{2}-h_{1}-h_{P}+1}z_{31}^{h_{1}-h_{2}-h_{P}+1}} \\ & \qquad \times\int_{\bar{z}_{2}}^{\bar{z}_{1}}\frac{d\bar{z}_{3} \ O_{(h_{P},\bar{h}_{P})}(z_{3},\bar{z}_{3})}{\bar{z}_{12}^{\bar{h}_{1}+\bar{h}_{2}+\bar{h}_{P}-1}\bar{z}_{32}^{\bar{h}_{2}-\bar{h}_{1}-\bar{h}_{P}+1}\bar{z}_{13}^{\bar{h}_{1}-\bar{h}_{2}-\bar{h}_{P}+1}}
\end{aligned}
\end{equation}
which has explicit $SL(2,\mathbb{R})_{L}\otimes SL(2,\mathbb{R})_{R}$
covariance. To recover the set of right descendants we perform an
expansion of the left factor to leading order in $z_{12}$ by setting
$z_{3}=z_{2}+tz_{12}$. We obtain  
\begin{align}
I_{O_{1}O_{2}}^{O_{3}} & =\frac{2\sin(\pi(h_{2}{-}h_{1}{+}h_{P}))}{z_{12}^{h_{1}+h_{2}-h_{P}}}\int_{1}^{\infty}dt\,t^{h_{1}+h_{P}-h_{2}-1}(t-1)^{h_{2}+h_{P}-h_{1}-1} \\
&\qquad \qquad \qquad \qquad \qquad \qquad \qquad \times \int_{\bar{z}_{2}}^{\bar{z}_{1}}\frac{d\bar{z}_{3} \ O_{(h_{P},\bar{h}_{P})}(z_{2}+tz_{12},\bar{z}_{3})}{\bar{z}_{12}^{\bar{h}_{1}+\bar{h}_{2}+\bar{h}_{P}-1}\bar{z}_{32}^{\bar{h}_{2}-\bar{h}_{1}-\bar{h}_{P}+1}\bar{z}_{13}^{\bar{h}_{1}-\bar{h}_{2}-\bar{h}_{P}+1}}\nonumber \\
 & =\frac{2\pi\Gamma(1{-}2h_{P})}{\Gamma(1{+}h_{2}{-}h_{1}{-}h_{P})\Gamma(1{+}h_{1}{-}h_{2}{-}h_{P})z_{12}^{h_{1}+h_{2}-h_{P}}}\int_{\bar{z}_{2}}^{\bar{z}_{1}}\frac{d\bar{z}_{3} \ O_{(h_{P},\bar{h}_{P})}(z_{2},\bar{z}_{3})}{\bar{z}_{12}^{\bar{h}_{1}+\bar{h}_{2}+\bar{h}_{P}-1}\bar{z}_{32}^{\bar{h}_{2}-\bar{h}_{1}-\bar{h}_{P}+1}\bar{z}_{13}^{\bar{h}_{1}-\bar{h}_{2}-\bar{h}_{P}+1}}\nonumber \\
 &\,\,\, \ \ +\mathcal{O}(z_{12}^{-h_{1}-h_{2}+h_{P}+1})\,.
\end{align}
We thus identify the remaining real integral as a conformal block for
$SL(2,\mathbb{R})_{R}$ descendants. It can be written in a more compact
way by introducing $\bar{z}_{3}=\bar{z}_{2}+t\bar{z}_{12}$: 
\begin{align}
SL(2,\mathbb{R})_{R}\textrm{ block} & =\int_{\bar{z}_{2}}^{\bar{z}_{1}}\frac{d\bar{z}_{3} \ O_{(h_{P},\bar{h}_{P})}(z_{2},\bar{z}_{3})}{\bar{z}_{12}^{\bar{h}_{1}+\bar{h}_{2}+\bar{h}_{P}-1}\bar{z}_{32}^{\bar{h}_{2}-\bar{h}_{1}-\bar{h}_{P}+1}\bar{z}_{13}^{\bar{h}_{1}-\bar{h}_{2}-\bar{h}_{P}+1}}\,\nonumber \\
 & =\frac{1}{\bar{z}_{12}^{\bar{h}_{2}+\bar{h}_{1}-\bar{h}_{P}}}\int_{0}^{1}\frac{dt \ O_{(h_{P},\bar{h}_{P})}(z_{2},\bar{z}_{2}+t\bar{z}_{12})}{t^{\bar{h}_{2}-\bar{h}_{1}-\bar{h}_{P}+1}(1-t)^{\bar{h}_{1}-\bar{h}_{2}-\bar{h}_{P}+1}}\,.
\end{align}
Let us apply this formula to the positive-helicity gluon case, for which $(h,\bar{h})=(\frac{\Delta+1}{2},\frac{\Delta-1}{2})$
and $\Delta_{P}=\Delta_{1}+\Delta_{2}-1$. We thus find 
\begin{equation}
I_{O_{1}O_{2}}^{O_{3},c}=\frac{2\pi\Gamma(1-\Delta_{1}-\Delta_{2})}{\Gamma(1-\Delta_{1})\Gamma(1-\Delta_{2})}\times\frac{1}{z_{12}}\int_{0}^{1}\frac{dt \ O_{\Delta_{P}}^{c,+}(z_{2},\bar{z}_{2}+t\bar{z}_{12})}{t^{2-\Delta_{1}}(1-t)^{2-\Delta_{2}}}+ \mathcal{O}(z_{12}^0)\,.
\end{equation}
The normalization factor $\mathcal{N}$ can be fixed in this case as  
\begin{align}
O_{\Delta_{1}}^{a,+}O_{\Delta_{2}}^{b,+} & \sim-if^{ab}{}_c\frac{\Gamma(1-\Delta_{1})\Gamma(1-\Delta_{2})}{2\pi\,\Gamma(1-\Delta_{1}-\Delta_{2})}I_{O_{1}O_{2}}^{O_{3},c}\nonumber \\
 & =\frac{-if^{ab}{}_c}{z_{12}}\int_{0}^{1}\frac{dt \ O_{\Delta_{P}}^{c,+}(z_{2},\bar{z}_{2}+t\bar{z}_{12})}{t^{2-\Delta_{1}}(1-t)^{2-\Delta_{2}}}\,,
\end{align}
in agreement with \eqref{eq:OOblock}.

\bibliography{softoperators.bib}
\bibliographystyle{utphys}

\end{document}